# High resolution and high precision beam profile monitor for particle accelerator using linear contact image sensor

Tung-Yuan Hsiao, Huan Niu, Tzung-Yuang Chen, and Chien-Hsu Chen[a]

*Accelerator Laboratory, Nuclear Science and Technology Development Center, National Tsing Hua University, Hsinchu, Taiwan*

[a]Author to whom correspondence should be addressed: akiracc@gmail.com, achchen@mx.nthu.edu.tw

## ABSTRACT

A compact beam-profile monitor was constructed using a linear contact image sensor attached to a plastic scintillator and tested using a 230 MeV proton beam. The results indicate that the beam distribution can be obtained in real-time, and the beam position with a precision of up to 0.03 mm. The compactness and high precision of the device hold considerable potential for it to be used as a beam-profile monitor and offline, daily quality assurance monitor in hadron therapy.

## I. INTRODUCTION

High-energy ion beams (proton and carbon) hold great potential in the field of radiation oncology due to their special Bragg peak characteristics, permitting the deposition of most of the beam's energy within a narrow depth range.[1] In radiation therapy, accurate and precise delivery of a specific radiation dose to the tumor site is crucial, and different beam monitors have been developed accordingly. These devices have been equipped with on beam-line for beam adjustment and also for day-to-day beam quality assurance. Ion chambers are the most commonly used devices to monitor the delivery of a dose from the therapy unit and comprise mainly two types: strip type and pixel matrix ionization chamber.[2-4] However, for more advanced treatments, such as hadron therapy, a faster and more precise beam-measurement device is needed. Therefore, plastic scintillators (PS) have been developed for verifying the proton range, to act as a quality assurance tool in clinical usage.[5]

The fast response of a plastic scintillator is essential for its usefulness as a suitable sensor for monitoring beam direction and position during the scan of an object. Moreover, the light yield of a plastic scintillator is dose-dependent, potentially allowing it to be used as a dose monitor. However, until now, the plastic scintillator system has required a relatively large space for its installation because it was observed by a long focal lenses camera, and the long focal lenses may cause image distortion. It is necessary to position the camera off the beam axis to prevent damage to it caused by high-energy particle bombardment.[5] These requirements have limited the practical application of plastic scintillators. The use of scintillator fibers can obviate the installation-space challenge, but it unavoidably presents a complex and expensive front-end readout system problem.[6]

The Linear Contact Image Sensor (CIS) module is the core device of a commercially available scanner.[7, 8] It consists of an array of microlenses to focus light onto the corresponding photodiode detector array, which converts the light to an electronic signal. The slim shape and ultrashort focus length of the CIS make it suitable for attachment to a plastic scintillator. The proton and radiation therapy center of Chang Gung Memorial Hospital is the first proton center in Taiwan, which has a cyclotron and four treatment rooms with a rotating gantry for proton therapy. Recently, an experimental room was set up for high-energy proton irradiation for physics and biology research, as shown in Figure 1. As a site for various types of experiment, a flexible, fast-response, and simply constructed beam monitoring system is required. We present the first-time use of a plastic scintillator attached to a slim CIS (PS+CIS) to monitor the position and profile of a high-energy proton beam.

## II. METHODS AND EXPERIMENT

A high-energy proton beam was produced using a cyclotron (P235, SHI, Japan). The beam was passed through a transporting line, becoming incident to the experimental room, passing through an exit window and then to the air, to where the test PS+CIS was placed on a bench. The inset of Figure 1 is a schematic diagram of the PS+CIS test system. Figure 2 shows the picture of the test configuration of the BPM in the CGMH experimental room. The PS was covered by a black sheet in case of the influence of indoor lighting. A 60-mm-thick copper block with a 10-mm-diameter hole was placed in front of the PS+CIS test system to function as a beam-shape collimator, and used to decrease the beam jitter. The PS is BC-408. The dimensions of the PS were $100 \times 100 \times 3$ mm$^3$, and the density 1.03 g/cm$^3$, which is similar to the density of tissue. The CIS used in this test was around 187 mm long with a resolution of 100 dpi (254 μm) at a scan speed < 20 μS/line, and was attached to the top-edge side of the PS. The plastic scintillator converted proton deposition energy into light, which was collected by the CIS and converted to electric signals, which were then fed to the front-end readout electronics. The readout electronics comprised a Cortex-M3 RISC processor with 10-bit ADC (analog to digital converter), which sent data to a PC by a Raspberry Pi Model 3 B (signal board computer), equipped with a 1.2-GHz 64-bit quad-core ARM Cortex-A53 and 1 G-Byte of LPDDR2 RAM. A EBT3 film,[9] was appended to the backside of the test PS+CIS, which recorded the beam information for comparison purposes.

## III. RESULTS AND DISCUSSION

Before the experiment, the background level of PS+CIS was determined in a dark box. Figure 3 shows the background spectrum of the test PS+CIS, which primarily originated from the CIS dark current; the background level was 80 units. Thus, the threshold level of PS+CIS was set at 80 and the acquired data shifted to 80 units. Moreover, the ADC readout limitation has been customized as 944 due to the maximum number of 10 bits ADC being 1024.

Figure 4(a) shows the measurement results for a proton-beam energy of 230 MeV and current of 1 nA, passing through the test PS+CIS for 21 s.

The response time of the PS and CIS are very short, and the measurement time is mainly limited by the followed analog-to-digital conversion time, the data transfer, and the data processing. For the sake of clarity, we designed the system so that one measurement took about 1 s; each measurement contained 726 data points and the total number of measurements was about 21. For human eye view, they were real-time, curve-fitted by a Gaussian equation, where the mean and standard deviation for the beam position and size were shown.

The statistical information of the fitted curves is shown in Figure 4(b), information obtained while the taking of a measurement was stopped or paused. Table 1 shows the fitting results and their FWHMs of the 21 measurements. The mean of peak position is 41.99 mm, and its standard deviation is 30 μm. These results indicate that the precision of the measured beam position is 30 μm. Figure 4(c) shows the scanned result of the irradiated EBT3 film, the comparison of which with the PS+CIS results, in the same projection direction, is shown in Figure 4(d). The coherent distributions between them indicate that the PS+CIS is a suitable beam profile monitor. Figure 5 shows the results of the PS+CIS at different beam intensities, without the beam-shape collimator.

Note that the light yield of the plastic scintillator is associated with the energy deposited on it by the radiation. The signal height increases with beam intensity up to a signal unit of 800, which could be caused by the saturation level of the photodiode. The signal shape broadens at high beam current, perhaps due to some stray light passing through the edges of the focus lens. The integration of the signal intensity of each measurement vs the correspondent beam current is shown in Figure 6. The linearity is more pronounced in the low-intensity region than in the high-intensity region, indicating that the dynamic range for 230 MeV protons in this system is around 0.5 nA to 2 nA.

## IV. CONCLUSION

In conclusion, we have presented a novel, high-resolution, and high-precision beam-profile monitor for particle accelerators, tested with a 230 MeV proton beam. The resolution of BPM is 254 μm which defined by the 100 dpi CIS sensor we used. And the results shows that the 100 dpi CIS sensor equipped with PS can achieve precision of beam position measurement up to 30 μm by 21 measurements, and the precision is significant better than found in traditional ion chamber type detectors. Due to the physical properties of the CIS+PS and their compact size, this beam profile monitor can be inserted into a beam line with limited space, and operated in real time as a beam-parameter monitor for high energy ion beam.

## ACKNOWLEDGMENTS

This work was financially supported by Team Union Ltd. We thank the Particle Physics and Beam Delivery Core Laboratory, Institute for Radiological Research, and the Chang Gung University/Chang Gung Memorial Hospital, Linkou, Taoyuan, Taiwan for their support in the experiments. We also thank Dr. Srinivasu Kunuku for his valuable input.

Table 1. The peak positions which were determined by Gaussian fit and their FWHM.

| No. | peak position | FHWM |
|---|---|---|
| 1 | 42.02 | 13.71 |
| 2 | 41.98 | 13.81 |
| 3 | 42.01 | 13.38 |
| 4 | 42.03 | 13.73 |
| 5 | 41.95 | 13.34 |
| 6 | 42.02 | 13.78 |
| 7 | 41.94 | 12.99 |
| 8 | 42.01 | 13.13 |
| 9 | 42.06 | 13.39 |
| 10 | 42.02 | 13.57 |
| 11 | 42.02 | 13.45 |
| 12 | 41.98 | 13.25 |
| 13 | 41.95 | 13.18 |
| 14 | 42.00 | 13.30 |
| 15 | 41.96 | 13.38 |
| 16 | 42.02 | 13.46 |
| 17 | 41.98 | 13.73 |
| 18 | 42.01 | 13.45 |
| 19 | 41.98 | 13.84 |
| 20 | 41.92 | 13.45 |
| 21 | 42.00 | 13.60 |
| Mean | 41.99 | 13.47 |
| Std | 0.03 | 0.24 |

Unit: mm

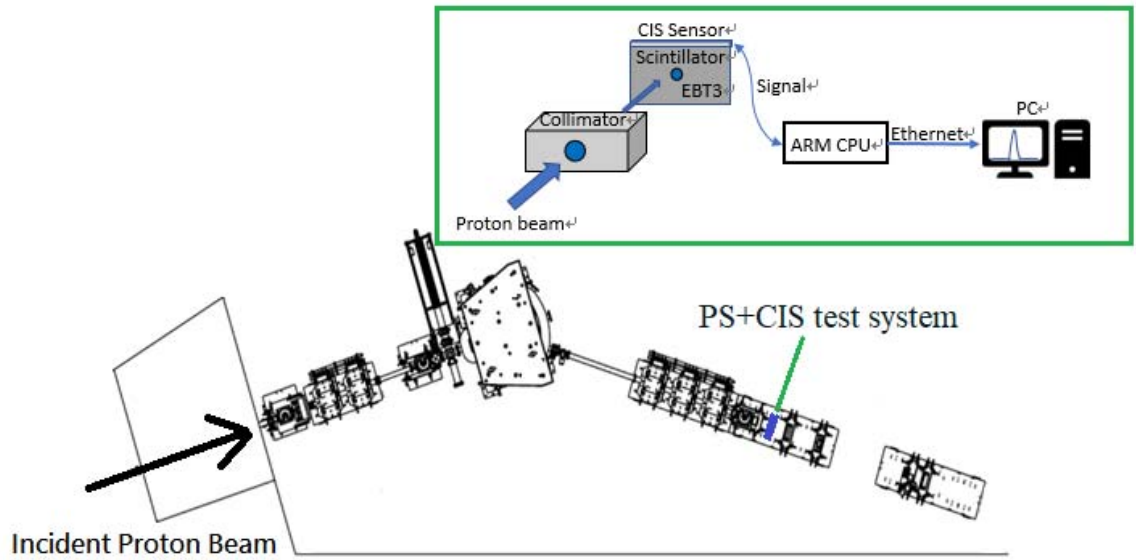

FIG. 1. Layout of proton irradiation experiment room in CGMH: inset shows schematic diagram of the PS+CIS test system.

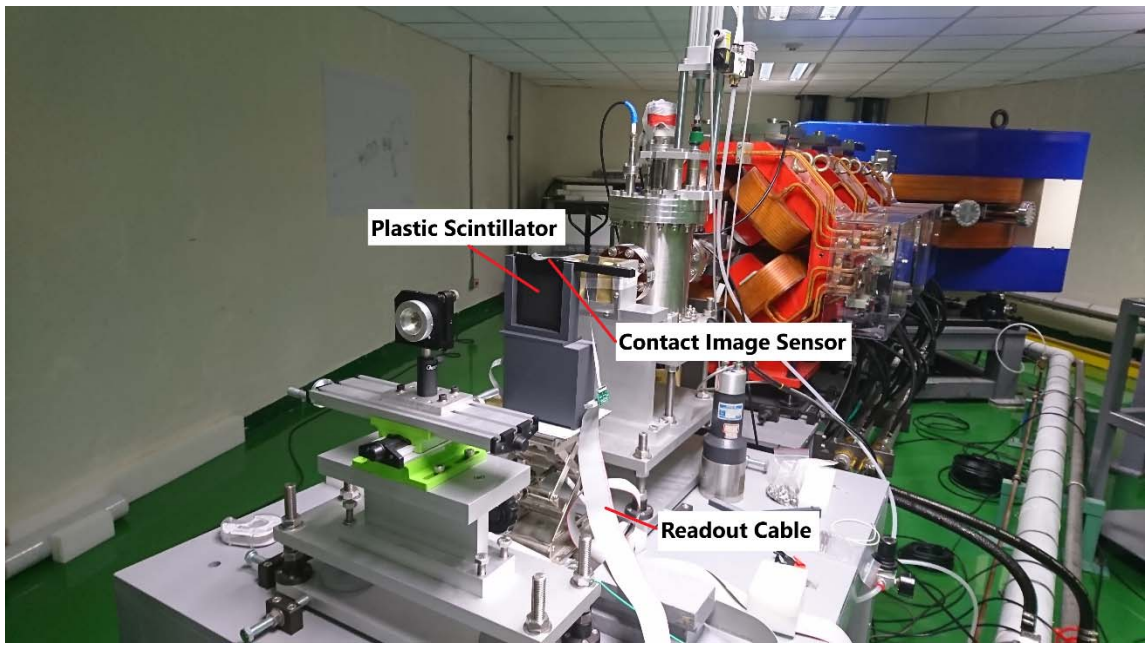

FIG. 2. The picture of the test configuration of the BPM in the CGMH experimental room.

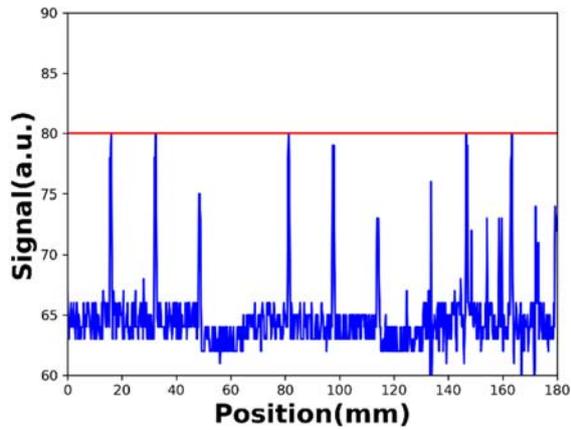

FIG. 3. Background spectrum obtained in a dark box; the threshold level (red line) is set at a value of 80 units.

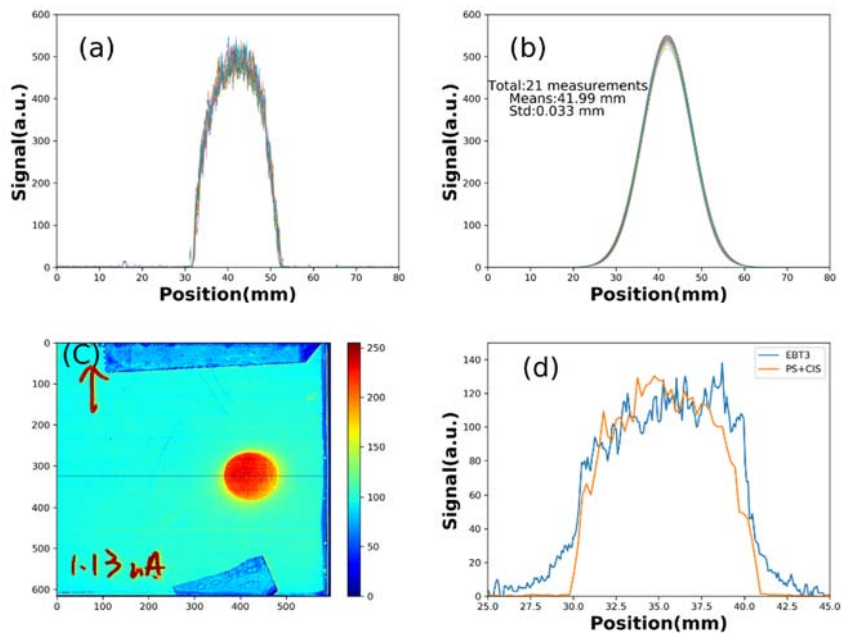

FIG. 4. Results of 230 MeV, 1 nA proton beam with a 1-cm collimator: (a) raw data of 21 measurements, (b) correspondent Gaussian fitting curves, (c) ETB3 film, (d) comparison between EBT3 and PS+CIS.

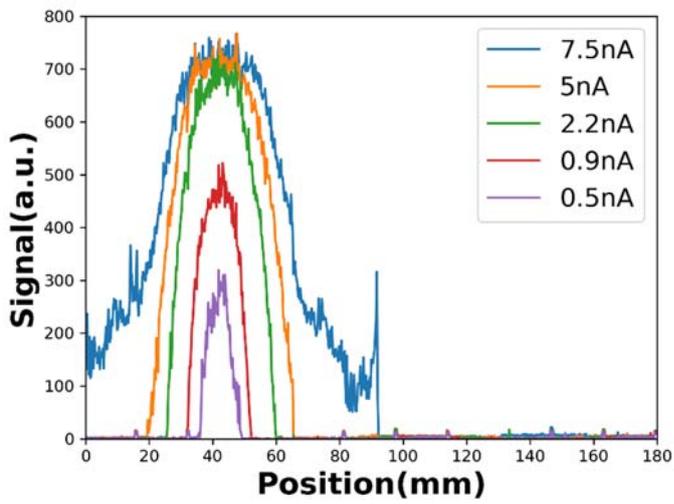

FIG. 5. Results of 230 MeV proton beam, without the collimator, at different currents.

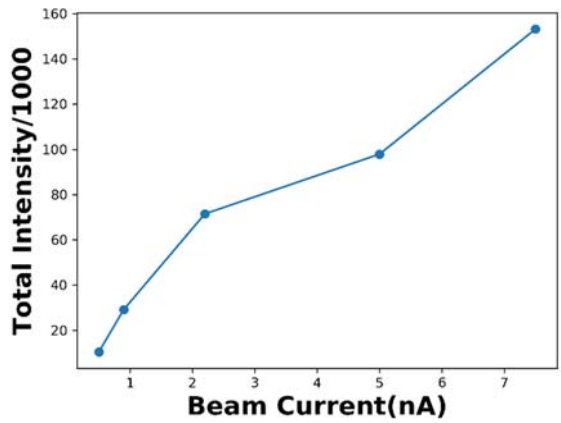

FIG. 6. The integration of the signal intensity vs correspondent beam current.